\documentclass[12pt]{article}
\usepackage{amsmath}
\usepackage{amsthm}
\usepackage{amssymb}
\usepackage{amscd}
\usepackage{amsfonts}
\usepackage{latexsym}
\input xy
\xyoption{all}

\newcommand{\bra}[1]{\ensuremath{\langle #1 |}}
\newcommand{\ket}[1]{\ensuremath{| #1 \rangle}}

\newcommand{\ot}{\otimes}
\newcommand{\ti}{\times}

\newcommand{\cD}{{\cal D}}

\newcommand{\cL}{{\cal L}}
\newcommand{\cJ}{{\cal J}}
\newcommand{\cS}{{\cal S}}

\newcommand{\cC}{{\cal C}}

\newcommand{\cH}{{\cal H}}

\newcommand{\la}{\langle}
\newcommand{\ra}{\rightarrow}
\newcommand{\ran}{\rangle}

\newcommand{\C}{{\mathbb C}}

\newcommand{\wt}{\widetilde}

\newcommand{\ol}{\overline}

\newcommand{\tr}{\mbox{$\text{Tr}$}}

\mathchardef\za="710B  
\mathchardef\zb="710C  
\mathchardef\zg="710D  
\mathchardef\zd="710E  
\mathchardef\zve="710F 
\mathchardef\zz="7110  
\mathchardef\zh="7111  
\mathchardef\zvy="7112 
\mathchardef\zi="7113  
\mathchardef\zk="7114  
\mathchardef\zl="7115  
\mathchardef\zm="7116  
\mathchardef\zn="7117  
\mathchardef\zx="7118  
\mathchardef\zp="7119  
\mathchardef\zr="711A  
\mathchardef\zs="711B  
\mathchardef\zt="711C  
\mathchardef\zu="711D  
\mathchardef\zvf="711E 
\mathchardef\zq="711F  
\mathchardef\zc="7120  
\mathchardef\zw="7121  
\mathchardef\ze="7122  
\mathchardef\zy="7123  
\mathchardef\zf="7124  
\mathchardef\zvr="7125 
\mathchardef\zvs="7126 
\mathchardef\zf="7127  
\mathchardef\zG="7000  
\mathchardef\zD="7001  
\mathchardef\zY="7002  
\mathchardef\zL="7003  
\mathchardef\zX="7004  
\mathchardef\zP="7005  
\mathchardef\zS="7006  
\mathchardef\zU="7007  
\mathchardef\zF="7008  
\mathchardef\zW="700A  

\newcommand{\be}{\begin{equation}}
\newcommand{\ee}{\end{equation}}

\newcommand{\bea}{\begin{eqnarray}}
\newcommand{\eea}{\end{eqnarray}}
\newcommand{\beas}{\begin{eqnarray*}}
\newcommand{\eeas}{\end{eqnarray*}}

\newtheorem{theorem}{Theorem}
\newtheorem{prop}{Proposition}

\newtheorem{cor}{Corollary}
\newtheorem{ex}{Example}
\newtheorem{definition}{Definition}

\newcommand{\epf}{\hfill$\Box$}
\newcommand{\bepf}{\noindent\textit{Proof.-} }

\begin{document}

\title{On the relation between states and maps\\ in infinite dimensions}

\author{Janusz Grabowski\footnote{email: jagrab@impan.gov.pl}  \\
\textit{Polish Academy of Sciences, Institute of Mathematics,} \\
\textit{ \'Sniadeckich 8, P.O. Box 21, 00-956 Warsaw, Poland}\\
\\
Marek Ku\'s\footnote{email: marek.kus@cft.edu.pl}  \\
\textit{Center for Theoretical Physics, Polish Academy of Sciences,} \\
\textit{Aleja Lotnik{\'o}w 32/46, 02-668 Warszawa, Poland}\\
\\
Giuseppe Marmo\footnote{email: marmo@na.infn.it}  \\
\textit{Dipartimento di Scienze Fisiche, Universit\`{a}
``Federico II'' di Napoli} \\
\textit{and Istituto Nazionale di Fisica Nucleare, Sezione di Napoli,} \\
\textit{Complesso Universitario di Monte Sant Angelo,} \\
\textit{Via Cintia, I-80126 Napoli, Italy}}

\maketitle

\begin{abstract}
Relations between states and maps, which are known for quantum systems in
finite-dimensional Hilbert spaces, are formulated rigorously in geometrical
terms with no use of coordinate (matrix) interpretation. In a tensor product
realization they are represented simply by a permutation of factors. This leads
to natural generalizations for infinite-dimensional Hilbert spaces and a simple
proof of a generalized Choi Theorem. The natural framework is based on spaces
of Hilbert-Schmidt operators $\cL_2(\cH_2,\cH_1)$ and the corresponding tensor
products $\cH_1\ot\cH_2^*$ of Hilbert spaces. It is proved that the
corresponding isomorphisms cannot be naturally extended to compact (or bounded)
operators, nor reduced to the trace-class operators. On the other hand, it is
proven that there is a natural continuous map
$\cC:\cL_1(\cL_2(\cH_2,\cH_1))\ra\cL_\infty(\cL(\cH_2),\cL_1(\cH_1))$ from
trace-class operators on $\cL_2(\cH_2,\cH_1)$ (with the nuclear norm) into
compact operators mapping the space of all bounded operators on $\cH_2$ into
trace class operators on $\cH_1$ (with the operator-norm). Also in the
infinite-dimensional context, the Schmidt measure of entanglement and
multipartite generalizations of state-maps relations are considered in the
paper.
\end{abstract}

\section{Introduction}
Entanglement is one of the most counterintuitive features of quantum mechanics.
Already in the early years of the birth of quantum theory, Erwin Schr\"odinger
realized that this aspect is a consequence of the mathematical structure of the
theory \cite{schroedinger35}. It is a characteristic property of quantum
mechanics, not present in other physical theories described by linear equations
like, for instance, classical electrodynamics. Entanglement was considered with
embarrassment in connection with non locality, as pointed out by the
Einstein-Podolsky-Rosen gedanken experiment \cite{einstein35}. Its role was
clarified with the discovery of Bell's inequalities \cite{bell64,bell66}. It
was shown that these inequalities can be violated in quantum mechanics but have
to be satisfied by all local realistic theories. The violation of the
inequalities demonstrates the presence of entanglement. In the near past it was
realized that entanglement could be a great resource for quantum information
theory. The promising applications of this peculiar quantum property has
induced intensive experimental efforts to build entangled quantum states and
major theoretical efforts to understand the mathematical structure of
entanglement \cite{bouwmeester00}. To put the present paper into perspective,
let us briefly consider how the problem arises.

In the Dirac-Schr\"odinger picture of quantum mechanics, one identifies the
carrier space of quantum evolution with a complex separable Hilbert space
$\mathcal{H}$. The probabilistic interpretation of quantum mechanics requires
that states be identified with rays, points of the complex projective space of
$\mathcal{H}$. By using the Hermitian inner product one defines an action of
the unitary group with an associated momentum map
\cite{grabowski05,grabowski06,carinena07}. This map relates rays with rank-one
projectors, i.e.\ operators, elements of the dual vector space of the Lie
algebra of the unitary group. Thanks to this immersion, it becomes possible to
consider convex combinations of elements in the image of this map and therefore
to construct density states, also called density operators. In this way
observables and states are represented both by means of operators, even though
with qualifying different properties to take into account their corresponding
physical interpretations. A similar situation results in the $C^*$-algebraic
approach to quantum mechanics, originated by Heisenberg and developed by Segal
and Haag \cite{segal47,haag92}. Here one considers states as nonnegative
normalized linear functionals on the space of observables, real elements of the
$C^*$-algebra and associates with them density states by means of Gleason's
theorem \cite{gleason57}.In either approach states are identified with
appropriate operators.

Composite systems are mathematically formed as tensor products of the Hilbert
spaces associated with the system we are composing, called subsystems.
Similarly within the $C^*$-algebra approach, the consideration of states as
maps has boosted a search for various procedures to characterize separability
and entanglement of states by exploiting as much as possible what is available
for the classification of maps \cite{sudarshan61,zyczkowski04,asorey05}.
The difficulties in a straightforward application of known classification
procedures rely on the fact that the very definition of states as convex
combination of rank-one projectors provides them with a positivity property
which is not preserved under tensorial products, in general the product of
positive maps does not result into a positive one. While the existing
literature is concerned with the relation between maps and states restricted to
finite dimensional Hilbert spaces or $C^*$-algebras, the aim of this paper is
to present a careful analysis of these various relations between states and
maps for composite quantum systems in the more realistic situation of infinite
dimensions.

Relations between states and maps are well known for systems in
finite-dimensional Hilbert spaces. In the second section of the paper we
reformulate them without invoking any particular matrix realization of the
states. This allows us to generalize in the following sections the known
results to infinite-dimensional Hilbert spaces and Hilbert-Schmidt operators
acting as maps between them. As a result we can describe in the
infinite-dimensional setting connections between positivity and complete
positivity of maps and separability properties of the corresponding states
on the composite spaces proven by Jamio{\l}kowski and Choi for the
finite-dimensional case. We discuss briefly generalization to multipartite
systems and show that the infinite-dimensional Jamio{\l}kowski isomorphism can
be neither sensibly extended to the larger class of bounded operators nor
reduced to a smaller set of the trace-class operators.

\section{The Jamio{\l}kowski isomorphism}
Let $\mathcal{H}_1$ and $\mathcal{H}_2$ be two Hilbert spaces. In a finite
dimensional case, $dim\mathcal{H}_{1,2}\le\infty$, the Jamio{\l}kowski
isomorphism \cite{jamiolkowski72,jamiolkowski75} is a mapping:
\begin{equation}\label{jam0}
\cJ:\mathcal{L}\left( \mathcal{L}\left( \mathcal{H}_{2}\right) ,
\mathcal{L}\left( \mathcal{H}_{1}\right) \right) \longrightarrow
\mathcal{L}\left( \mathcal{L}\left( \mathcal{H}_{2}, \mathcal{H}_{1}\right)
\right),
\end{equation}
where by $\mathcal{L}\left( \mathcal{H}_{2}, \mathcal{H}_{1}\right)$ we denote
the space of all complex linear maps from $\mathcal{H}_2$ to $\mathcal{H}_1$,
with an abbreviation $\mathcal{L}\left( \mathcal{H}\right)=\mathcal{L}\left(
\mathcal{H},\mathcal{H}\right)$ for the space of all linear endomorphisms of
$\mathcal{H}$.

We prefer to define objects in a basis-independent way, so we prefer to speak
about $\cH$ and $\cL(\cH)$ instead of $\C^n$ and complex matrices. Therefore we
start with the following observations. First, there is a natural
anti-isomorphism between $\mathcal{H}$ and its dual complex vector space,
\begin{equation}\label{1}
\zk_\cH:\mathcal{H}\rightarrow \mathcal{H}^{\ast },
\end{equation}
induced by the scalar product on $\mathcal{H}$, which in the Dirac notation
reads
\begin{equation*}
\mathcal{H}\ni x:=\left| x\right\rangle \mapsto \overline{x}:=\left\langle
x\right|\in\mathcal{H}^{\ast }.
\end{equation*}
The dual space $\mathcal{H}^\ast$ is canonically a Hilbert space with the
Hermitian product $\langle
\overline{x}_1,\overline{x}_2\rangle_{\cH^\ast}=\langle
x_2,x_1\rangle_{\cH}\,$, where the latter is the scalar product on
$\mathcal{H}$, which shows that $\zk$ is an anti-unitary. In the following we
usually skip the subscripts specifying the Hermitian products in various
spaces, if this does not lead to a confusion.

We clearly have $\zk_{\cH^\ast}\circ\zk_\cH=id_\cH$ up to an obvious
identification $(\cH^*)^*=\cH$. Moreover, the anti-isomorphism (\ref{1})
induces an anti-isomorphism
\begin{equation}\label{2}
\cL(\cH)\ni A\mapsto \ol{A}\in
\cL(\cH^\ast)
\end{equation}
of the corresponding spaces of complex linear operators, where
$\ol{A}=(A^\dag)^*$. Here, clearly, the adjoint operator $A^\dag\in\cL(\cH)$ is
defined by $\langle A^\dagger x,y\rangle_{\,\mathcal{H}_1}=\langle x, A
y\rangle_{\,\mathcal{H}_1}$ and $A^*\in\cL(\cH^*)$ is the dual map. By
definition, (\ref{1}) intertwines $A$ with $\ol{A}$, i.e.
$\ol{A(x)}=\ol{A}(\ol{x})$. The notation is consistent, because
$\ol{A}=\zk_{\cL(\cH)}(A)$ (up to an obvious identification
$\cL(\cH)^*\simeq\cL(\cH^*)$) for the Hermitian product $\left\langle {A,B}
\right\rangle  = {\rm{Tr}}\left( {A^\dagger\circ B} \right)$ on
$\mathcal{L}\left(\cH\right)$.

The point here is that $A\mapsto\ol{A}$ respects the composition, $\ol{A\circ
B}=\ol{A}\circ\ol{B}$, while $(AB)^*=B^*A^*$. This means that, restricting
ourselves to the groups of invertible complex operators on the Hilbert spaces,
we have a canonical group isomorphism $GL(\cH)\ni A\mapsto\ol{A}\in GL(\cH^*)$,
while $GL(\cH)\ni A\mapsto{A^*}\in GL(\cH^*)$ is an anti-isomorphism. This
group isomorphism restricts to an isomorphism of the unitary groups $U(\cH)\ni
A\mapsto\ol{A}\in U(\cH^*)$, as in this case
$$\la\ol{A}(\ol{x}),\ol{A}(\ol{y})\ran_{\cH^*}=\la Ay,Ax\ran_\cH=\la
y,x\ran_\cH=\la\ol{x},\ol{y}\ran_{\cH^*}\,.$$

Note that in the physics literature one usually identifies $\cH$ with
$\cH^{\ast}$ by fixing an orthonormal basis $(e_i)$ in $\cH$ and putting
$$\sum_ic_i\ket{e_i}\simeq\sum_ic_i\bra{e_i}\,.$$
It is a true isomorphism which, however, depends on the choice of the basis,
and not the canonical anti-isomorphism we speak about.

\medskip
We will use the following canonical identification of Hilbert spaces,
\begin{equation}\label{iso}
\mathcal{L}\left( \mathcal{H}_2,\mathcal{H}_1\right) =\mathcal{H}_1\otimes
\mathcal{H}_2^{\ast }.
\end{equation}
Under this identification
$(x\otimes\overline{y})\left(y^\prime\right)=\left\langle y,
y^\prime\right\rangle x$ for $x\in\mathcal{H}_1$ and
$y,y^\prime\in\mathcal{H}_2$. Moreover, the Hilbert-Schmidt scalar product
$\left\langle {A,B} \right\rangle  = {\rm{Tr}}\left( {A^\dagger\circ B}
\right)$ on $\mathcal{L}\left( \mathcal{H}_2,\mathcal{H}_1\right)$ coincides
with the standard scalar product $\left\langle {x_1  \otimes \overline y _1
,x_2 \otimes \overline y_2 } \right\rangle  = \left\langle {x_1 ,x_2 }
\right\rangle_{\mathcal{H}_1}\left\langle {y_2 ,y_1 }
\right\rangle_{\mathcal{H}_2}$ on $\mathcal{H}_1\otimes \mathcal{H}_2^{\ast }$.
Here, the adjoint operator
\begin{equation}
A^{\dagger }\in \mathcal{L}\left( \mathcal{H}_{1},\mathcal{H}_{2}\right)
=\mathcal{H} _{2}\otimes \mathcal{H}_{1}^{\ast }
\end{equation}
is defined in an obvious way by $\langle A^\dagger
x,y\rangle_{\,\mathcal{H}_2}=\langle x, A y\rangle_{\,\mathcal{H}_1}$ (or, in
the tensor product realization, $(x\ot\ol{y})^\dag=y\ot\ol{x}$). Indeed, for
$\{f_\alpha\}$ being an arbitrary orthonormal basis in $\mathcal{H}$,
\begin{eqnarray*}
{\rm{Tr}}\big( {\left({x_1 \otimes \overline{y}_1} \right)^\dagger\circ
\left({x_2 \otimes \overline{y}_2} \right)} \big) &=& {\rm{Tr}}\left(
{\left\langle {x_1 ,x_2 } \right\rangle y_1 \otimes \overline{y}_2} \right) =
 \left\langle{x_1,x_2}\right\rangle\sum\limits_\alpha
 {\left\langle{f_\alpha,\left( {y_1  \otimes \overline{y}_2 }
 \right)f_\alpha}\right\rangle}
 = \\
 &=&\left\langle {x_1 ,x_2 } \right\rangle \sum\limits_\alpha
 {\left\langle {f_\alpha  ,y_1 } \right\rangle }
 \left\langle {y_2 ,f_\alpha } \right\rangle
  = \left\langle {x_1 ,x_2 } \right\rangle \left\langle {y_2 ,y_1 }
  \right\rangle.
\end{eqnarray*}
We have a canonical $GL(\cH_1)\ti GL(\cH_2)$-action on $\mathcal{L}\left(
\mathcal{H}_{2},\mathcal{H}_{1}\right)$:
\begin{equation}\label{action}
GL(\cH_1)\ti GL(\cH_2)\ti\mathcal{L}\left(
\mathcal{H}_{2},\mathcal{H}_{1}\right) \ni (A,B,T)\mapsto A\circ T\circ
B^\dag\in\mathcal{L}\left( \mathcal{H}_{2},\mathcal{H}_{1}\right)
\end{equation}
which in
the tensor product realization takes the form
\begin{equation}\label{action1}
GL(\cH_1)\ti
GL(\cH_2)\ti\cH_1\ot\cH_2^* \ni (A,B,x\ot\ol{y})\mapsto A(x)\ot
\ol{B}(\ol{y})\in\cH_1\ot\cH_2^*\,.
\end{equation}
This action can be reduced to an $U(\cH_1)\ti U(\cH_2)$-action which is
unitary, as in this case
$$\la Ax\ot\ol{By},Ax'\ot\ol{By'}\ran=\la Ax,Ax'\ran\la\ol{By},\ol{By'}
\ran=\la x,x'\ran\la y',y\ran=\la x\ot\ol{y},x'\ot\ol{y'}\ran\,.$$
If $\cH_1=\cH_2$, then one can reduce the above action to a diagonal
action of $GL(\cH)$ (or $U(\cH)$): $(A,T)\mapsto A\circ T\circ A^\dag $.

The canonical isomorphism $\cH_1\ot\cH_2^\ast\simeq\cH_2^\ast\ot\cH_1$ gives
rise to an identification
\begin{equation}\label{id1}
\mathcal{L}\left(\mathcal{H}_2,\mathcal{H}_1\right) \simeq\mathcal{L}\left(
\mathcal{H}_1^\ast,\mathcal{H}_2^\ast\right)\,. \end{equation}
Moreover,
\begin{equation}\label{id2}
\mathcal{L}\left( \mathcal{H}_2,\mathcal{H}_1\right)^\ast
\simeq\mathcal{L}\left( \mathcal{H}_1,\mathcal{H}_2\right)\,,
\end{equation}
with the obvious pairing
\begin{equation}\label{id3}
\mathcal{L}\left(\mathcal{H}_2,\mathcal{H}_1\right) \ti\mathcal{L}\left(
\mathcal{H}_1,\mathcal{H}_2\right)\,\ni (A,B)\mapsto \tr(A\circ B)\in\C.
\end{equation} In particular,
\begin{equation}\label{id4}
\mathcal{L}\left(\mathcal{H}\right)^\ast\simeq\mathcal{L}\left(
\mathcal{H}\right)\simeq\cL\left(\cH^*\right)\,.
\end{equation}
Note that we have further natural identifications
$$\mathcal{L}\left( \mathcal{L}\left( \mathcal{H}_{2}\right) ,
\mathcal{L}\left( \mathcal{H}_{1}\right) \right)=\mathcal{H}_{1}\otimes
\mathcal{H}_{1}^{\ast }\otimes \mathcal{H}_{2}\otimes \mathcal{H}_{2}^{\ast }$$
and
$$\mathcal{L}\left( \mathcal{L}\left( \mathcal{H}_{2},
\mathcal{H}_{1}\right) \right)=\mathcal{H}_{1}\otimes\mathcal{H}_{2}^{\ast
}\otimes \big(\mathcal{H}_{1}\otimes \mathcal{H}_{2}^{\ast }\big)^{\ast}=
\mathcal{H}_{1}\otimes\mathcal{H}_{2}^{\ast }\otimes \mathcal{H}_{2}\otimes
\mathcal{H}_{1}^{\ast }.$$

\begin{definition}
{The {\em Jamio\l kowski isomorphism}
(\ref{jam0}), up to above natural identifications, is defined as a natural
transposition in the tensor products
\begin{equation}\label{J0}
\cJ:\mathcal{H}_{1}\otimes \mathcal{H}_{1}^{\ast }\otimes
\mathcal{H}_{2}\otimes \mathcal{H}_{2}^{\ast }\longrightarrow
\mathcal{H}_{1}\otimes\mathcal{H}_{2}^{\ast }\otimes \mathcal{H}_{2}\otimes
\mathcal{H}_{1}^{\ast }
\end{equation}
consisting of interchanging of the second and fourth factors, i.e.\
\begin{equation}\label{J}
\cJ:x_{1}\otimes \overline{x}_{2}\otimes y_{1}\otimes \overline{y}_{2}\mapsto
x_{1}\otimes \overline{y}_{2}\otimes y_{1}\otimes \overline{x}_{2}.
\end{equation}
The {\it twisted Jamio\l kowski isomorphism}
\begin{equation}\label{twisted}
\wt{\cJ}:\mathcal{L}\left( \mathcal{L}\left(
\mathcal{H}_{2}\right),\cL\left(\cH_1\right)\right)\ra\cL(\cH_1\ot\cH_2)
\end{equation}
comes in a similar way from the permutation
\bea\label{J0tw}
\wt{\cJ}:\mathcal{H}_{1}\otimes \mathcal{H}_{1}^{\ast }\otimes
\mathcal{H}_{2}\otimes \mathcal{H}_{2}^{\ast }&\longrightarrow &
\mathcal{H}_{1}\otimes\mathcal{H}_{2}\otimes \mathcal{H}_{2}^{\ast }
\otimes\mathcal{H}_{1}^{\ast }\\
x_{1}\otimes \overline{x}_{2}\otimes y_{1}\otimes \overline{y}_{2}&\mapsto
&x_{1}\otimes y_{1}\otimes\overline{y}_{2}\otimes \overline{x}_{2}\,. \eea}
\end{definition}
As Jamio\l kowski isomorphisms are simply permutations in the tensor product,
they are automatically unitary. Moreover, it is completely clear that Jamio\l
kowski isomorphisms intertwine the canonical actions of the group $GL(\cH_1)\ti
GL(\cH_1)\ti GL(\cH_2)\ti GL(\cH_2)$ on the tensor product
$\mathcal{H}_{1}\otimes \mathcal{H}_{1}^{\ast }\otimes \mathcal{H}_{2}\otimes
\mathcal{H}_{2}^{\ast }$:
\begin{equation}\label{actionj} (A,A',B,B',x_{1}\otimes
\overline{x}_{2}\otimes y_{1}\otimes \overline{y}_{2})\mapsto Ax_{1}\otimes
\overline{A'x}_{2}\otimes By_{1}\otimes \overline{B'y}_{2}
\end{equation} and its
corresponding permutations, so the following is immediate.
\begin{theorem}\label{intertwi}
The Jamio\l kowski isomorphisms are unitary and intertwine the canonical
$GL(\cH_1)\ti GL(\cH_1)\ti GL(\cH_2)\ti GL(\cH_2)$-actions.
\end{theorem}

\section{Infinite dimensions}
\medskip
The above definitions can be extended to infinite-dimensional  Hilbert spaces
$\mathcal{H}_1$ and $\mathcal{H}_2$ as follows. In this case we define the
Hilbert-Schmidt tensor product $\mathcal{H}_1\overline{\otimes}\,\mathcal{H}_2$
as the closure of the algebraic tensor product
$\mathcal{H}_1\otimes_\mathbb{C}\mathcal{H}_2$ with respect to the scalar
product which on simple tensors reads $\langle x\otimes y, x^\prime\otimes
y^\prime\rangle=\langle x, x^\prime\rangle_{\mathcal{H}_1}\langle y,
y^\prime\rangle_{\mathcal{H}_2}$. In this way, elements of
$\mathcal{H}_1\overline{\otimes}\,\mathcal{H}_2$ represent Hilbert-Schmidt
operators from $\cH_2^{\ast}$ into $\cH_1$ and can be viewed as infinite
combinations $A=\sum_{i,\alpha }a_{i\alpha }\,(e_{i}\otimes f_{\alpha })$,
where $(e_i)$ and $f_\za$ are orthonormal bases in $\cH_1$ and $\cH_2$,
respectively, and $\Vert A\Vert_2^2=\sum_{i,\alpha }\left| a_{i\alpha }\right|
^{2}<\infty $ is the squared Hilbert-Schmidt norm of $A$ (it does not depend on
the choice of bases). It is well-known \cite{reed72} that every compact operator
$A:\cH_2\ra\cH_1$ admits the so called {\it Schmidt decomposition},
$A=\sum_j\zl_j\la a_j,\cdot\ran b_j\,$, with $(a_j)$ and $(b_j)$ being (not
necessarily complete) orthonormal sets, and $\zl_j{\ra}0$ as $j\ra\infty$. The
Hilbert-Schmidt norm can be equivalently defined as $\Vert
A\Vert_2^2=\sum_j\vert\zl_j\vert^2$. In fact, the coefficients $\zl_j$ can be
chosen positive. In the following we will denote
$\mathcal{H}_1\overline{\otimes}\,\mathcal{H}_2$ simply as
$\mathcal{H}_1\otimes\mathcal{H}_2$. Since the Hermitian conjugation is also a
transposition of the tensor product, $A^\dag$ is Hilbert-Schmidt if $A$ is.

\medskip
Now, the (Hillbert-Schmidt) tensor product $\cH_1\ot\cH_2^{\ast}$ represents
the space $\mathcal{L}_2(\mathcal{H}_2,\mathcal{H}_1)$ of  the Hilbert-Schmidt
operators, i.e.\ the Hilbert space of those complex  linear maps $A$ from
$\mathcal{H}_2$ to $\mathcal{H}_1$ such that $\sum_{i}\left\langle
Af_{\alpha},Af_{\alpha}\right\rangle_{\mathcal{H}_1}<\infty $, for some (thus
all) orthonormal basis $\{f_{\alpha}\}$ in $\mathcal{H}_2$, and with the
Hermitian form
$$\la{A},{B}\ran=\tr(A^\dag B)=\sum_{i}\left\langle
Af_{\alpha},Bf_{\alpha}\right\rangle_{\mathcal{H}_1}.$$
Note that the trace is well-defined, since any composition of Hilbert-Schmidt
operators is known to be a {\it trace-class operator} \cite{reed72}.

We will abbreviate $\mathcal{L}_2(\mathcal{H},\mathcal{H})$ to
$\mathcal{L}_2(\mathcal{H})$ for an arbitrary Hilbert space $\mathcal{H}$. The
symbol $\cL$ is now reserved for all bounded complex linear maps, so that
$\cL(\cH_2,\cH_1)$ is the space of all bounded operators from $\cH_2$ to
$\cH_1$ with the operator-norm topology. The latter makes sense also in the
Banach category.

In complete analogy with the finite-dimensional case we have also natural
canonical identifications
$$\mathcal{L}_2\left( \mathcal{L}_2\left( \mathcal{H}_{2}\right) ,
\mathcal{L}_2\left( \mathcal{H}_{1}\right) \right)=\left(\mathcal{H}_{1}
\otimes \mathcal{H}_{1}^{\ast }\right)\otimes\left( \mathcal{H}_{2}\otimes
\mathcal{H}_{2}^{\ast }\right)^\ast=\mathcal{H}_{1}\otimes
\mathcal{H}_{1}^{\ast }\otimes\mathcal{H}_{2}\otimes \mathcal{H}_{2}^{\ast }$$
and
$$\mathcal{L}_2\left( \mathcal{L}_2\left( \mathcal{H}_{2}, \mathcal{H}_{1}
\right) \right)=\left(\mathcal{H}_{1}\otimes\mathcal{H}_{2}^{\ast
}\right)\otimes \big(\mathcal{H}_{1}\otimes \mathcal{H}_{2}^{\ast
}\big)^{\ast}= \mathcal{H}_{1}\otimes\mathcal{H}_{2}^{\ast }\otimes
\mathcal{H}_{2}\otimes \mathcal{H}_{1}^{\ast }.$$
Thus, the Jamio{\l}kowski isomorphism, defined on the level of tensor products
by the same transposition (\ref{J}), is now interpreted as
\begin{equation}\label{jamhs}
\cJ:\mathcal{L}_2\big(\mathcal{L}_2(\mathcal{H}_2),
\mathcal{L}_2(\mathcal{H}_1)\big) {\rightarrow
}\mathcal{L}_2\big(\mathcal{L}_2(\mathcal{H}_2,\mathcal{H}_1)\big),
\end{equation}
and the twisted Jamio\l kowski isomorphism, as
\begin{equation}\label{twisted2}
\wt{\cJ}:\mathcal{L}_2\left( \mathcal{L}_2\left(
\mathcal{H}_{2}\right),\cL_2\left(\cH_1\right)\right)\ra\cL_2(\cH_1\ot\cH_2)\,.
\end{equation}
Both isomorphisms are clearly unitary. Moreover, since the Hilbert-Schmidt
operators form an operator ideal, the Hilbert-Schmidt tensor products are
invariant with respect to the canonical $GL(\cH_1)\ti GL(\cH_1)\ti GL(\cH_2)\ti
GL(\cH_2)$-actions and the Jamio\l kowski isomorphisms intertwines these
actions.

Note also that the original definitions are obviously equivalent to the
following properties of the Jamio\l kowski isomorphisms.
\begin{prop}
{The Jamio\l kowski unitary isomorphisms (\ref{jamhs}) and (\ref{twisted2})
can be uniquely characterized, respectively, by the identities
\begin{equation}\label{Jsp} \left\langle x\otimes\overline{y},
\cJ(\Phi)\left(x^\prime\otimes\overline{y^\prime}\right)\right\rangle=
\left\langle x\otimes\overline{x^\prime},
\Phi\left(y\otimes\overline{y^\prime}\right)\right\rangle
\end{equation}
and
\begin{equation}\label{Jsp2}
\left\langle x\otimes{y},
\wt{\cJ}(\Phi)\left(x^\prime\otimes{y}'\right)\right\rangle= \left\langle
x\otimes\overline{x^\prime},
\Phi\left(y'\otimes\overline{y}\right)\right\rangle,
\end{equation}
which must be fulfilled for all
$x,x^\prime\in\mathcal{H}_1$, $y,y^\prime\in\mathcal{H}_2$. Equivalent
formulations of the above identities are, respectively,
\begin{equation}\label{Jsp'} \left\langle
y\ot\ol{x}\ot x^\prime\otimes\overline{y^\prime},\cJ(\Phi)\right\rangle=
\left\langle x'\ot\ol{x}\ot y\otimes\overline{y^\prime}, \Phi\right\rangle
\end{equation}
and
\begin{equation}\label{Jsp2'}
\left\langle\ol{y}\ot\ol{x}\ot x^\prime\otimes{y}',
\wt{\cJ}(\Phi)\right\rangle= \left\langle x'\ot\ol{x}\ot y'\otimes\overline{y},
\Phi \right\rangle\,.
\end{equation}}
\end{prop}
\begin{ex}
{\rm
For $\rho \in \mathcal{L}_2\left( \mathcal{H}_{2}\right)$, orthonormal bases
$(x_l)$ and $(y_\za)$ in $\cH_1$ and $\cH_2$, and for
\begin{equation}
A,B\in \mathcal{L}_2\left( \mathcal{H}_{2},\mathcal{H}_{1}\right)
=\mathcal{H}_{1}\otimes \mathcal{H}_{2}^{\ast },\quad
A=\sum_{l\za}A^{l\alpha}\cdot x_{l}\otimes\overline{y}_{\alpha},\quad
B=\sum_{l,\za}B^{l\alpha}\cdot x_{l}\otimes\overline{y}_{\alpha},
\end{equation}
let us consider the map
\begin{equation}\label{L}
M_A^B\in \mathcal{L}_2\left( \mathcal{L}_2\left( \mathcal{H}_{2}\right)
,\mathcal{L}_2\left( \mathcal{H} _{1}\right) \right) ,\quad M_A^B:\rho \mapsto
A\rho B^{\dagger },
\end{equation}
i.e.\
\begin{equation}
M_A^B:y_{\alpha}\otimes \overline{y}_{\beta}\mapsto
\sum_{l,m,\za,\zb}A^{l\alpha}\overline{B^{m\beta}}\cdot x_{l}\otimes
\overline{x}_{m}.
\end{equation}
Hence, the Jamio\l kowski isomorphism applied to $M_A^B$ is a map
$\cJ(M_A^B)\in\cL_2(\mathcal{L}_2\left(
\mathcal{H}_{2},\mathcal{H}_{1}\right))$ represented by
\begin{eqnarray}\label{LA}
\cJ(M_A^B)&=&\cJ(\sum_{l,m,\za,\zb}A^{l\alpha}\overline{B^{m\beta}}\cdot
x_{l}\otimes \overline{x}_{m}\otimes y_{\beta}\otimes
\overline{y}_{\alpha})\\
&=&\sum_{l,m,\za,\zb}A^{l\alpha}\overline{B^{m\beta}}\cdot
x_{l}\otimes \overline{y}_{\alpha}\otimes y_{\beta}\otimes\overline{x}_{m} \\
\nonumber &=&\sum_{l,m,\za,\zb}\left( A^{l\alpha}\cdot x_{l}\otimes
\overline{y}_{\alpha}\right) \otimes\overline{\left( B^{m\beta}\cdot
x_{m}\otimes \overline{y}_{\beta}\right) } =A\otimes \ol{B},
\end{eqnarray}
i.e.\ $\cJ(M_A^B)$ is just the one-dimensional operator $\ket{A}\bra{B}$. In
particular, if $A=B$, the operator $M_A^A=K_A$ is just the standard Kraus map
$K_A(\zr)=A\zr A^\dag$ and its Jamio\l kowski image $\cJ(K_A)$ is the Hermitian
"projection" $p_A=\ket{A}\bra{A}$ on the vector $A\in\mathcal{L}_2\left(
\mathcal{H}_{2},\mathcal{H}_{1}\right)$. This is a true projection if the
length of $A$ is 1. The fact that we deal with a unitary isomorphism implies
easily that there is an orthonormal basis in $\mathcal{L}_2\left(
\mathcal{L}_2\left( \mathcal{H}_{2}\right) ,\mathcal{L}_2\left( \mathcal{H}
_{1}\right) \right)$ consisting of operators of the form $M_{A_j}^{A_k}$ for a
basis $(A_j)$ in $\mathcal{L}_2\left( \mathcal{H}_{2},\mathcal{H}_{1}\right)$.

Since $\cJ$ is unitary, the Hilbert-Schmidt norm of $K_A$ equals the
Hilbert-Schmidt norm of this projection, i.e. $\Vert A\Vert_2^2$. Recall that
the space of Hilbert-Schmidt operators is an operator ideal in the space of all
bounded operators as $\Vert A\circ\zr\Vert_2\le \Vert
A\Vert_\infty\Vert\zr\Vert_2$, where $\Vert\cdot\Vert_\infty$ is the
operator-norm. Since $\Vert K_A(\zr)\Vert_2\le\Vert
A\Vert_\infty^2\Vert\zr\Vert_2$, the operator-norm $\Vert K_A\Vert_\infty$ of
$K_A$ is not bigger than the square of the operator-norm of $A$, i.e. $\Vert
K_A\Vert_\infty\le\Vert A\Vert_\infty^2$. But the operator-norm of the
projection $A\ot\ol{A}$ is still $\Vert A\Vert_2^2$. Since we can easily find a
sequence $(A_n)$ with $\Vert A_n\Vert_2=1$ such that $\Vert A_n\Vert_\infty\ra
0$, this shows that the Jamio\l kowski isomorphism is not continuous in the
operator-norm topology. In other words, $\cJ$ does not admit a natural
extension to a map
$$\cJ:\mathcal{L}_\infty\big(\mathcal{L}_2(\mathcal{H}_2),
\mathcal{L}_2(\mathcal{H}_1)\big) {\rightarrow
}\mathcal{L}_\infty\big(\mathcal{L}_2(\mathcal{H}_2,\mathcal{H}_1)\big)\,, $$
where $\cL_\infty$ denotes the space of compact operators -- the operator-norm
closure of the space of all Hilbert-Schmidt operators. One can think that the
above suggests that the inverse Jamio\l kowski isomorphism $\cJ^{-1}$ is
continuous in the norm topology, as
$$\Vert J^{-1}(p_A)\Vert_\infty=\Vert K_A\Vert_\infty\le
\Vert p_A\Vert_2.$$ But this is also not true, since
$\Vert\sum_1^np_{A_i}\Vert_\infty$ is 1 for $A_i=e\ot f_i$, $\Vert e\Vert=1$,
and the operator-norm of the corresponding Kraus map
$$\cL_2(\cH_2)\ni\zr\mapsto\sum_1^nA_i\zr A_i^\dag\in\cL_2(\cH_1)$$
is at lest $\sqrt{n}$. Indeed, the projection $P_n$ on the subspace spanned by
$f_1,\dots,f_n$ has the Hilbert-Schmidt norm $\sqrt{n}$, while its image
$$\sum_1^nA_iP A_i^\dag=\sum_1^n\la f_i,Pf_i\ran e\ot\ol{e}$$
has the Hilbert-Schmidt norm $n$.}
\end{ex}

Any composition of Hilbert-Schmidt operators is well known to be a {\it
trace-class operator}, called also {\it nuclear operator} (see e.g.
\cite[Chapter VII]{Maurin72}). The space of nuclear operators $T:\cH_2\ra\cH_1$
consists of operators admitting a decomposition into one-dimensional operators:
$Tx=\sum_i\la a_i,x\ran b_i$ (in $\cL_2(\cH_2,\cH_1)=\cH_1\ot\cH_2^\ast$ they
are represented as tensors that can be written in the form
$\sum_ib_i\ot\ol{a_i}$) with $\sum_i\Vert a_i\Vert\!\cdot\!\Vert
b_i\Vert<\infty$. They can be equivalently described as these operators for
which $T^\dag T$ is Hilbert-Schmidt on $\cH_2$. The nuclear norm can be defined
as $\Vert T\Vert_1=\sum_\za{\Vert Tf_\za\Vert}=\tr\left(\sqrt{TT^\dag}\right)$,
as $\Vert T\Vert_1=\sum_i\zl_i$ for any Schmidt decomposition $T=\sum_i\zl_i\la
a_i,\cdot\ran b_i$ with $(a_i)$ and $(b_i)$ being (not necessarily complete)
orthonormal sets, or as the {\it infimum} of $\sum_i\Vert
a_i\Vert\!\cdot\!\Vert b_i\Vert$ for all possible realizations. The latter has
the advantage that it applies also in the Banach space context.
\begin{ex}
{\rm For an orthonormal base $(x_l)$ in $\cH_1$, a vector $y\in\cH_2$ of length
1, and for a sequence of complex numbers $a=(a_l)\in l^2$, the operator
$T:\cH_2\ra\cH_1$,
$$T=\left\ket{\sum_l a_l\cdot x_l\ot\ol{x_l}\right}\bra{y\ot\ol{y}}=
\sum_la_l\cdot x_l\ot\ol{x_l}\ot{y\ot\ol{y}}\,,$$ is nuclear with the nuclear
norm $\Vert T\Vert_1=\Vert a\Vert_2=\sqrt{\sum_l\vert a_l\vert^2}$. Its Jamio\l
kowski image is Hermitian
$$\cJ(T)=\sum_la_l\ket{x_l\ot\ol{y}}\bra{x_l\ot\ol{y}}=\sum_la_l
\cdot x_l\ot\ol{y}\ot y\ot\ol{x_l}$$ with eigenvalues $(a_l)$, so
$\Vert\cJ(T)\Vert_1=\Vert a\Vert_1=\sum_l\vert a_l\vert$. Since there are
sequences $a\in l^2$ with infinite $l^1$-norm, the Jamio\l kowski isomorphism
does not map nuclear operators into nuclear ones. A similar fact can be proved
for the inverse Jamio\l kowski isomorphism.}
\end{ex}
Let us summarize the
conclusions of the above examples in the following proposition.
\begin{prop}
The Jamio\l kowski isomorphism (\ref{jamhs}) and its inverse
cannot be extended to all compact operators nor restricted to nuclear
(trace-class) operators.
\end{prop}
Despite of the above negative result, it is obvious that the map
$M_A^B:\zr\mapsto A\zr B^\dag$, associated with $A,B\in\cL_2(\cH_2,\cH_1)$, can
be viewed as a map $M_A^B:\cL(\cH_2)\ra\cL_1(\cH_1)$.
\begin{theorem}
 There is a unique continuous map
\begin{equation}\label{choi-map}
\cC:\cL_1(\cL_2(\cH_2,\cH_1))\ra\cL_\infty(\cL(\cH_2),\cL_1(\cH_1))
\end{equation}
such that $\cC(A\ot\ol{B})(\zr)=A\zr B^\dag$ for all $A,B\in\cL_2(\cH_2,\cH_1)$.
\end{theorem}
\bepf Let us start with computing the operator-norm of $M_A^B$. First of all,
for an orthonormal basis $(e_j)$ in $\cH_1$,  we have $\zr B^\dag x=\sum_j\la
e_j,\zr B^\dag x\ran e_j$, so
$$A\zr B^\dag x=\sum_j\la e_j,\zr B^\dag x\ran Ae_j=\sum_j
\la B\zr^\dag e_j, x\ran Ae_j\,.$$
Hence,
$$\Vert A\zr B^\dag\Vert_1\le\sum_j\Vert B\zr^\dag e_j\Vert\cdot\Vert Ae_j
\Vert\le\left(\sum_j\Vert B\zr^\dag
e_j\Vert^2\right)^{\frac{1}{2}}\left(\sum_j\Vert
Ae_j\Vert^2\right)^{\frac{1}{2}}=\Vert B\zr^\dag\Vert_2 \Vert A\Vert_2\,.
$$
But, as easily seen, $\Vert B\zr^\dag\Vert_2\le\Vert B\Vert_2\Vert
\zr\Vert_\infty$, so $\Vert A\zr B^\dag\Vert_1\le\Vert A\Vert_2\Vert
B\Vert_2\Vert\zr\Vert_\infty$ and $\Vert M_A^B\Vert_\infty\le\Vert
A\Vert_2\Vert B\Vert_2=\Vert A\ot\ol{B}\Vert_1$. Moreover, if $(A_j)$ is an
orthonormal basis in $\in\cL_2(\cH_2,\cH_1)$, then
$$\Vert\sum_{j,k}\zl_j^kM_{A_k}^{A_j}\Vert_\infty\le\sum_{j,k}
\vert\zl_j^k\vert\Vert
M_{A_k}^{A_j}\Vert_\infty\le\sum_{j,k}\vert\zl_j^k\vert=\Vert
\sum_{j,k}\zl_j^kA_k\ot\ol{A_j}\Vert_1$$
which shows that $\cC$ is bounded (continuous) with the operator-norm $\le 1$.
One can easily see that this norm is actually 1. Let us see that the operators
$M_A^B$ are compact. Indeed, \beas
M_A^B&=&\sum_{l,m,\za,\zb}A^{l\alpha}\overline{B^{m\beta}}\cdot
x_{l}\otimes \overline{x}_{m}\ot y_\zb\ot\ol{y_\za}\\
&=&\sum_{l+\za\le N}\sum_{m,\zb} A^{l\alpha}\overline{B^{m\beta}}\cdot
x_{l}\otimes \overline{x}_{m}\ot y_\zb\ot\ol{y_\za}+ \sum_{l+\za>
N}\sum_{m,\zb}A^{l\alpha}\overline{B^{m\beta}}\cdot x_{l}\otimes
\overline{x}_{m}\ot y_\zb\ot\ol{y_\za}\,. \eeas But the operator
$$\sum_{l+\za\le N}\sum_{m,\zb} A^{l\alpha}\overline{B^{m\beta}}\cdot
x_{l}\otimes \overline{x}_{m}\ot y_\zb\ot\ol{y_\za}$$ is finite-dimensional and
$$R_N=\sum_{l+\za> N}\sum_{m,\zb} A^{l\alpha}\overline{B^{m\beta}}\cdot
x_{l}\otimes \overline{x}_{m}\ot y_\zb\ot\ol{y_\za}$$ has the operator-norm
$$\Vert R_N\Vert_\infty\le\left(\sum_{l+\za>N}\vert
A^{l\za}\vert^2\right)^\frac{1}{2}\Vert B\Vert_2$$
which is arbitrary small for large $N$, so $M_A^B$ is an operator-norm limit of
finite-dimensional operators. Further, $\sum_{j,k}\zl_j^kM_{A_k}^{B_j}$ is
clearly compact for $A_k,B_j$ of length 1 if
$\sum_{j,k}\vert\zl_j^k\vert<\infty$.

\epf

\noindent The operator $\cC$ we will call the {\it Choi map}.

\medskip
The trace-class operators appear in Quantum Mechanics as quantum states. The
convex set $\cD(\cH)$ of {\it quantum states} consists of trace-class
non-negative Hermitian operators with trace 1. It follows from the spectral
theorem that each quantum state $\zr$ can be written in a form
$\zr=\sum_i\zl_i\zr_i$, where $(\zr_i)$ is a sequence of one-dimensional
orthogonal projectors, with $\la\zr_i,\zr_j\ran=0$ for $i\ne j$, and $\zl_i\ge
0$, $\sum_i\zl_i=1$. In other words, $\cD(\cH)$ is the smallest convex set in
$\cL(\cH)$ closed in the nuclear topology which contains all pure states --
one-dimensional orthogonal projectors. The Choi map associates with quantum
states $\sum_j\zl_j\ket{A_j}\bra{A_j}$, with $\Vert A_j\Vert_2=1$, on
$\cL_2(\cH_2,\cH_1)$ a Kraus maps from $\cL(\cL(\cH_2),\cL_1(\cH_1))$ with the
operator sum representation $\zr\mapsto\sum_j\zl_jA_j\zr A_j^\dag$.

If $\cH=\cH_1\ot\cH_2$, a quantum state $\zr\in\cD(\cH)$ we call {\it
separable} if one can find a decomposition $\zr=\sum_i\zl_i\zr_i$ as above but
with $\zr_i$ being simple tensors, $\zr_i=\zr_i'\ot\zr_i''$, where $\zr_i'$ and
$\zr_i''$ are one-dimensional orthogonal projectors in, respectively, $\cH_1$
and $\cH_2$. Composite quantum states we call {\it entangled} if they are not
separable. Replacing $\cH_2$ with $\cH_2^*$ we can speak, in an obvious sense,
about separable and entangled quantum states on $\cL_2(\cH_2,\cH_1)$.

\section{Basic features of $\cJ$}
To proceed we shall need some further observations. First, let us see that maps
from a linear subspace $V$ of $\cL(\cH_2)$, closed with respect to the
Hermitian conjugation, into $\cL(\cH_1)$, which preserve hermiticity, commute
with the operation of taking the adjoint. Indeed, assume that
$\Phi:\cL(\mathcal{H}_2)\supset V\ra\cL(\mathcal{H}_1)$ is a linear map that
maps Hermitian operators into Hermitian ones, and define, for an arbitrary
$A\in V$,
\begin{equation*}
\Psi(A)=\Phi(A)^\dagger-\Phi(A^\dagger).
\end{equation*}
Clearly, $\Psi$ is additive, $\Psi(A+B)=\Psi(A)+\Psi(B)$, and antilinear,
$\Psi(\alpha A)=\overline{\alpha}\,\Psi(A)$. Now,
\begin{equation}\label{ApA}
\Psi(A+A^\dagger)=\Phi\big(A+A^\dagger\big)^\dagger- \Phi\big((A+A^\dagger)^\dagger\big)=
\Phi\big(A+A^\dagger\big)-\Phi\big(A+A^\dagger\big)=0,
\end{equation}
where we use the hermiticity of $A+A^\dagger$ and the assumed property that
$\Phi$ maps Hermitian operators into Hermitian ones. Since $A$ was arbitrary,
we can take $iA$ instead of $A$, hence
\begin{equation*}
0=\Psi\big(iA+(iA)^\dagger\big)=\Psi\big(iA-iA^\dagger\big)=
-i\Psi\big(A-A^\dagger\big),
\end{equation*}
which, upon additivity of $\Psi$ and together with (\ref{ApA}) gives
$\Psi(A)=0$. Thus we get the following proposition.
\begin{prop}
If \ $V\subset\cL(\mathcal{H}_2)$ is a linear subspace,
closed with respect to Hermitian conjugation, and $\Phi:V\ra\cL(\mathcal{H}_1)$
is a linear map that maps Hermitian operators into Hermitian ones, then $\Phi$
commutes with Hermitian conjugation, $\Phi(A^\dag)=\Phi(A)^\dag$.
\end{prop}
\begin{theorem}
A Hilbert-Schmidt operator
$\Phi:\cL_2(\mathcal{H}_2)\ra\cL_2(\mathcal{H}_1)$ preserves hermiticity  if
and only if $\cJ(\Phi)$ is Hermitian.
\end{theorem}
\bepf According to the above proposition, preserving hermiticity means
commuting with the Hermitian conjugation. Since, fixing orthonormal bases
$(x_j)$ and $(y_a)$ in $\cH_1$ and $\cH_2$, respectively,
$(y_a\ot\ol{y_b})^\dag=y_b\ot\ol{y_a}$, etc.,
$\zF=\sum_{i,j,a,b}\zl_{ijab}x_i\ot\ol{x_j}\ot y_a\ot\ol{y_b}$ commutes with
the Hermitian conjugation if and only if $\zl_{ijab}=\ol{\zl_{jiba}}$. On the
other hand, $\cJ(\zF)$ is Hermitian if and only if
\beas\cJ(\zF)&=&\sum_{i,j,a,b}\zl_{ijab}\cdot x_i\ot\ol{y_b}\ot y_a\ot\ol{x_j}
= \left(\sum_{i,j,a,b}\zl_{ijab}\cdot x_i\ot\ol{y_b}\ot y_a\ot\ol{x_j}\right)^\dag\\
&=& \sum_{i,j,a,b}\ol{\zl_{ijab}}\cdot x_j\ot\ol{y_a}\ot y_b\ot\ol{x_i}\,,
\eeas
i.e., as above, if and only if $\zl_{ijab}=\ol{\zl_{jiba}}$.

\epf

We say that $\Phi$ as above {\it preserves positivity} (this property is
usually called also {\it positivity} that might be confused with positivity of
a Hermitian operator), if it maps non-negatively defined Hermitian operators
on $\mathcal{H}_2$ (we will call them simply {\it positive}) into positive ones
on $\mathcal{H}_1$. Using (\ref{Jsp}) we can prove now the following.
\begin{theorem}
A Hilbert-Schmidt operator $\Phi:\cL_2(\mathcal{H}_2)\ra\cL_2(\mathcal{H}_1)$
preserves positivity if and only if $\cJ(\Phi)$ is a Hermitian operator on
$\cL_2(\cH_2,\cH_1)$ which is non-negatively defined on separable states, i.e.
$\tr(\cJ(\Phi)\zr)\ge 0$ for separable states $\zr$ on $\cL_2(\cH_2,\cH_1)$.
\end{theorem}
\bepf We have to prove that $\Phi$ preserves the positivity if and only if, for
arbitrary $x\in\mathcal{H}_1$, $y\in\mathcal{H}_2$,
\begin{equation}\label{positive}
\left\langle x\otimes\overline{y},\cJ(\Phi)\left(x\otimes\overline{y}\right)\right\rangle\ge 0.
\end{equation}
Indeed, assume that (\ref{positive}) holds. Then from (\ref{Jsp})
\begin{equation}\label{positive1}
\left\langle x\otimes\overline{x},\Phi\left(y\otimes\overline{y}\right)\right\rangle \ge 0.
\end{equation}
Hence, for each projection $x\otimes\overline{x}$ its Hilbert-Schmidt scalar
product with $\Phi$ evaluated on arbitrary $y\otimes\overline{y}$ is positive,
so $\Phi(y\ot\ol{y})$ is positive for all $y$ and then positivity of $\Phi(A)$
for arbitrary positive-definite $A$ follows from the spectral decomposition of
$A$.

On the other hand, if $\Phi$ preserves positivity, then evaluated on a positive
operator $y\otimes\overline{y}$ it gives a positive operator for which the
Hilbert-Schmidt scalar product with an arbitrary projection
$x\otimes\overline{x}$ is non-negative, hence (\ref{positive1}) and \emph{a
fortiori} (\ref{positive}) hold. \epf

A natural question now is: what Hilbert-Schmidt operators
$\Phi:\cL_2(\mathcal{H}_2)\ra\cL_2(\mathcal{H}_1)$ correspond, {\it via} the
Jamio\l kowski isomorphism, to Hermitian operators which are  positive on the
whole $\cL_2(\cH_2,\cH_1)$.
\begin{definition}
{A Hilbert-Schmidt operator $\Phi:\cL_2(\mathcal{H}_2)
\ra\cL_2(\mathcal{H}_1)$ we call {\em completely positive}, if $\cJ(\Phi)$ is
Hermitian positive on $\cL_2(\cH_2,\cH_1)$.}
\end{definition}

We will show now that the above natural definition is equivalent to the
standard concepts of complete positivity. Note however that we cannot consider
tensor products with the identity on an infinite-dimensional Hilbert space, as
the latter is not a Hilbert-Schmidt operator. Therefore, for an auxiliary
Hilbert space $\mathcal{H}$ with an orthonormal basis $(u_i)$, consider the
Hilbert-Schmidt operator $K_A$ on $\mathcal{L}_2(\mathcal{H})$ associated with
a diagonal Hilbert-Schmidt matrix $A=\sum_j\zl_j\cdot u_j\ot\ol{u_j}$,
$\sum_j\vert\zl_j\vert^2<\infty$. In other words,
\begin{equation}\label{Lambda}
K_A =\sum_{i,j}\lambda _{i}\overline{\lambda}_{j}\,u_{i}\otimes
\overline{u}_{j} \otimes u_{j}\otimes \overline{u}_{i}\,.
\end{equation}
We know that $\cJ(K_A)$ is Hermitian positive,
\begin{equation}\label{JLambda}
\cJ(K_A) =A\ot\ol{A}=\sum_{i,j}\lambda _{i}\overline{\lambda}_{j}\,u_{i}\otimes
\overline{u}_{i} \otimes u_{j}\otimes \overline{u}_{j}\, .
\end{equation}
For $\Phi\in\cL_2(\cL_2(\cH_2),\cL_2(\cH_1))$, we can consider $\Phi\ot
K_A\in\cL_2(\cL_2(\cH_2'),\cL_2(\cH_1'))$ with $\cH_i'=\cH_i\ot\cH$, $i=1,2\,$,
with its Jamio\l kowski image $\cJ(\Phi\ot K_A)\in\cL_2(\cL_2(\cH_2',\cH_1'))$.

Take now arbitrarily chosen $x_1,\dots,x_m\in\cH_1$, $y_1,\dots,y_m\in\cH_2$.
We have
\bea
\nonumber &&\left\langle\left(\sum_{k=1}^{m}x_{k}\otimes
u_{k}\right)\ot\ol{\left(\sum_{k=1}^{m}x_{k}\otimes u_{k}\right)},\left(\Phi\ot
K_A\right)\left(\sum_{k=1}^{m}y_{k}\otimes
u_{k}\right)\ot\ol{\left(\sum_{k=1}^{m}y_{k}\otimes u_{k}\right)}\right\rangle\\
&=&\left\langle \sum_{k}x_{k}\otimes \overline{\zl_ky}_{k},\cJ\left( \Phi
\right) \left( \sum_{p}x_{p}\otimes \overline{\zl_py}_{p}\right) \right\rangle.
\label{completely}\end{eqnarray}
Indeed, for
$$
X=\sum_{k=1}^{m}x_{k}\otimes u_{k}\in \mathcal{H}_{1}\otimes \mathcal{H}, \quad
Y=\sum_{k=1}^{m}y_{k}\otimes u_{k}\in \mathcal{H}_{2}\otimes \mathcal{H},
$$
one can write
\begin{eqnarray*}
&\phantom{=}&\left\langle X\otimes \overline{X},\left( \Phi \otimes \Lambda
\right) \left( Y\otimes \overline{Y}\right) \right\rangle=\left\langle X\otimes
\overline{Y},\cJ\left( \Phi \otimes \Lambda \right) \left( X\otimes
\overline{Y}\right) \right\rangle \\
&=&\sum_{k,l,p,q}^{m}\left\langle x_{k}\otimes \overline{y}_{l},\cJ\left( \Phi
\right)\left( x_{p}\otimes \overline{y}_{q}\right)\right\rangle \left\langle
u_{k}\otimes \overline{u}_{l},\cJ\left( \Lambda \right)
\left(u_{p}\otimes\overline{u}_{q}\right)\right\rangle \\
&=&\sum_{k,l,p,q}^{m}\left\langle x_{k}\otimes \overline{y}_{l},\cJ\left( \Phi
\right) x_{p}\otimes\overline{y}_{q}\right\rangle \left\langle u_{k}\otimes
\overline{u}_{l},(A\ot\ol{A})\left( u_{p}\otimes \overline{u}_{q}\right)
\right\rangle \\
&=&\sum_{k,l,p,q}^{m}\left\langle x_{k}\otimes \overline{y}_{l}, \cJ\left( \Phi
\right) x_{p}\otimes \overline{y}_{q}\right\rangle \sum_{i,j}\lambda
_{i}\overline{\lambda}_{j}\delta _{p}^{j}\delta _{q}^{j}\delta _{k}^{i}\delta
_{l}^{i} \\
&=&\sum_{k,p}^{m}\lambda _{k}\overline{\lambda}_{p}\left\langle x_{k} \otimes
\overline{y}_{k},\cJ\left( \Phi \right) x_{p}\otimes
\overline{y}_{p}\right\rangle  \\
&=&\left\langle \sum_{k}x_{k}\otimes \overline{\zl_ky}_{k},\cJ\left( \Phi
\right) \left( \sum_{p}x_{p}\otimes \overline{\zl_py}_{p}\right) \right\rangle.
\end{eqnarray*}
Note that any vector in $\cH_1\ot\cH_2^\ast$, thus any map in
$\cL_2(\cH_2,\cH_1)$, can be approximated by vectors of the form
$Z=\sum_{p}x_{p}\otimes \overline{\zl_py}_{p}$. Similarly, vectors from
$\mathcal{H}_{1}\otimes \mathcal{H}$ and $\mathcal{H}_{2}\otimes \mathcal{H}$
can be approximated by vectors of the form $X$ and $Y$ as above. If $\cH_i$,
$i=1,2$, are finite-dimensional, then we can actually get all these vectors
taking the number of $u_i$ not exceeding the maximum of these dimensions.
Since, according to the formula (\ref{completely}), $\langle X\otimes
\overline{X},\left( \Phi \otimes \Lambda \right) \left( Y\otimes
\overline{Y}\right)\ran\ge 0$ if and only if $\langle Z,\cJ\left( \Phi \right)
\left( Z\right) \rangle\ge 0$, we can derive the following characterization of
complete positivity that can be viewed as an infinite-dimensional version of
Choi Theorem \cite{choi75}, cf. also \cite{salgado05}.
\begin{theorem}
Let $\Phi\in\mathcal{L}_2\big(\mathcal{L}_2(\mathcal{H}_2),
\mathcal{L}_2(\mathcal{H}_1)\big)$.
The following are equivalent:
\begin{description}
\item{(a)} $\cJ(\Phi)$ is Hermitian positive;
\item{(b)} For any finite-dimensional Hilbert space $\cH$ the
operator $\Phi\ot I\in\cL_2(\cL_2(\cH_2\ot\cH),\cL_2(\cH_1\ot\cH))$ preserves
positivity;
\item{(c)} For an infinite-dimensional Hilbert space $\cH$ and for a Hermitian
positive $A\in\cL_2(\cH_2,\cH_1)$ with trivial kernel, the operator $\Phi\ot
K_A\in\cL_2(\cL_2(\cH_2\ot\cH),\cL_2(\cH_1\ot\cH))$ preserves positivity.
\end{description}
If $\cH_i$, $i=1,2$, are finite-dimensional, then the dimensions of above
auxiliary Hilbert spaces $\cH$ can be restricted to the maximum of the
dimensions of $\cH_1$ and $\cH_2$.
\end{theorem}
Of course, all the above has the corresponding counterpart for the other
Jamio\l kowski isomorphism $\wt{\cJ}$. This version fits sometimes better to
the language of bi-partite systems.
\section{Schmidt rank and Schmidt measure}
We know already that any element $v\in\cH_1\ot\cH_2$ admits a Schmidt
decomposition $\zf=\sum_j\zl_j\cdot a_j\ot b_j$ with $(a_j)$ and $(b_j)$ being
(not necessarily complete) orthonormal sets, and $\zl_j$ being positive. The
Hilbert-Schmidt norm can be equivalently defined as $\Vert
A\Vert_2^2=\sum_j\vert\zl_j\vert^2$. The number of summands in this
decomposition (which can be infinite if both Hilbert spaces are
infinite-dimensional) we call the {\it Schmidt rank\ } $\cS(\zf)$ of $\zf$.
Directly by definition, a pure state $p_\zf=\ket{\zf}\bra{\zf}$ on
$\cH_1\ot\cH_2$ is {\it separable} if and only if the Schmidt rank of $\zf$ is
1.
\begin{prop}
The Schmidt rank of $\zf$ is $m=1,2,\dots,\infty$ if and only
if $p_\zf=\wt{\cJ}(\Phi)$ for an operator $\Phi:\cL_2(\cH_2)\ra\cL_2(\cH_1)$ of
rank $m^2$.
\end{prop}
\bepf Since $\Phi=\sum_{j,k}\zl_j\zl_k\cdot a_j\ot\ol{a_k}\ot b_j\ot\ol{b_k}$
and $\zl_j\zl_k>0$ for $j,k=1,\dots,m$, \ the image of $\Phi$ is spanned by
$a_j\ot\ol{a_k}$, $j,k=1,\dots,m$, thus its rank is $m^2$.

\epf

\medskip\noindent This suggests the following extension of the concept of Schmidt rank.
\begin{definition}
{The {\em Schmidt rank} of $\zr\in\cL_2(\cH_1\ot\cH_2)$ is
the operator-rank of $\wt{\cJ}^{-1}(\zr)$.}
\end{definition}
In these terms we can state the following corollary, where we admit
infinite-dimensional Hilbert spaces.
\begin{cor}
A pure state $\zr$ on $\cH_1\ot\cH_2$
is separable if and only if the Schmidt rank of $\zr$ is 1.
\end{cor}
This easy characterization of separable pure states has been used by Terhal and
Horodecki \cite{terhal00} to develop the concept of {\it Schmidt number}  of an
arbitrary density matrix $\zr$ (quantum state in finite dimensions). This
number characterizes the minimum Schmidt rank of the pure states that are
needed to construct such density matrix. The Schmidt number is non-increasing
under local operations and classical communications, i.e.\ it provides a
legitimate entanglement measure. We can construct an entanglement measure --
{\it Schmidt measure $\zm_S$} -- which is additionally convex using the convex
roof construction (see e.g.\ \cite{eisert01}). This construction, proposed as a
general tool for entanglement measures (see e.g.\
\cite{uhlmann00,grabowski05,grabowski06}), can be repeated in infinite dimensions
as
\begin{equation}\label{S}
\zm_S(\zr)=\inf \left\{\sum_j\zl_j\cS(\zf_j)\right\}\,,
\end{equation}
where the {\it infimum} is taken over all possible realizations of $\zr$ as
infinite-convex combinations $\zr=\sum_j\zl_j\ket{\zf_j}\bra{\zf_j}$ with
$0\le\zl_j\le 1$, $\sum_j\zl_j=1$ and $\zf_j\in\cH_1\ot\cH_2$. Every quantum
state admits such a realization and a reasoning analogous to the one in
\cite{grabowski05} shows that $\zm_S$ is infinite-convex, non-negative and
vanishes exactly on separable states.
\section{Multipartite generalizations}
The diagram of the Jamio\l kowski isomorphisms

\begin{equation}\label{dam}
\xymatrix{
\cH_2\ot\cH_2^*\ot\cH_1\ot\cH_1^* \ar[ddr]_{\cJ}
 \ar[rr]_{\wt{\cJ}}
    && \cH_2\ot\cH_1^*\ot\cH_1\ot\cH_2^*\ar@{<->}[ddl]
 \\ \\
    & \cH_1\ot\cH_2\ot\cH_2^*\ot\cH_1^* &
}
\end{equation}
interpreted also as
\begin{equation}\label{dam1}
\xymatrix{
\cL_2(\cL_2(\cH_1),\cL_2(\cH_2)) \ar[ddrr]_{\cJ}
 \ar[rrr]_{\wt{\cJ}}
 & &  & \cL_2(\cL_2(\cH_1,\cH_2))\ar@{<->}[ddl]
 \\ \\
   & & \cL_2(\cH_1\ot\cH_2) & & }
\end{equation}
can be easily generalized to multipartite
cases, where we replace $\cH_1\ot\cH_2$ with $\cH_1\ot\cdots\ot\cH_n$. Of
course, the number of possible permutations grows quickly with $n$. Part of
them can be obtained by a trivial induction. For instance, we can replace
$\cH_2$ with $\cH_2\ot\cH_3$ (or $\cL_2(\cH_2,\cH_3)$) in (\ref{dam1}), but we
can also get
$$ \cL_2\left(\cL_2(\cL_2(\cH_1,\cH_2)\right),\cL_2(\cH_3))\longleftrightarrow
\cL_2(\cL_2(\cH_1\ot\cH_2),\cL_2(\cH_3))\longleftrightarrow
\cL_2(\cL_2(\cH_1,\cH_2\ot\cH_3))$$ or

$$\xymatrix{ \cL_2(\cL_2(\cH_1\ot\cH_2),\cL_2(\cH_3\ot\cH_4))
\ar@{<->}[dd]
 \ar@{<->}[r]
    & \cL_2(\cL_2(\cH_1\ot\cH_3),\cL_2(\cH_2\ot\cH_4))\ar@{<->}[dd]
 \\ \\
  \cL_2(\cL_2(\cL(\cH_1,\cH_4),\cH_3)\cL_2(\cH_2))\ar@{<->}[r]  &
  \cL_2(\cH_1\ot\cH_2\ot\cH_3\ot\cH_4)}$$
etc. We will not study here these isomorphisms in details, as the choice of a
particular one depends on our interests in possible applications.

\section{Acknowledgements}
This work was supported by the Polish Ministry of Scientific and Higher
Education under the (solicited) grant No PBZ-Min-008/P03/03, EU IP ``SCALA''
and partially supported by PRIN SINTESI.


\end{document}